\documentclass[a4paper,11pt]{article}
\pdfoutput=1 	
\usepackage{jheppub} 	
\usepackage{verbatim}
\usepackage{graphicx}	
\usepackage{dcolumn}	
\usepackage{bm}		    
\usepackage{hyperref}
\usepackage{amsmath}
\usepackage{verbatim}
\usepackage{slashed}
\usepackage{hyperref}
\usepackage{mathtools}
\usepackage{tabularx}

\title{\boldmath  Dynamical origin of Type-I Seesaw with large mixing}

\author{Yi Chung}

\affiliation{
Max-Planck-Institut für Kernphysik, Saupfercheckweg 1, 69117 Heidelberg, Germany
}

\emailAdd{yi.chung@mpi-hd.mpg.de}

\abstract{We investigate Type-I Seesaw models where the right-handed neutrino masses are dynamically generated by strong interactions. Using horizontal gauge symmetry as the source of strong dynamics, a nontrivial flavor structure can also be introduced dynamically. We find that the right-handed neutrino mass matrix with a strongly anti-diagonal structure emerges when the three right-handed neutrinos are in the triplet representation of $SU(2)_H$ horizontal gauge symmetry. With an assumption of a Dirac neutrino mass matrix with hierarchical eigenvalues and small mixing angles, analogous to the up-type quark sector, and certain substructures, the resulting light neutrino mass matrix from the type-I seesaw mechanism can accommodate the large mixing and weak hierarchy observed in low-energy neutrino data. The neutrino puzzles can, therefore, be understood as the consequence of strong horizontal gauge interactions. We also discuss the potential UV completion and the phenomenology that could be tested in the future.}

\begin{document}
\maketitle
\flushbottom

\section{Motivation}


The Standard Model (SM) of particle physics successfully describes all known elementary particles and their interactions. However, several puzzles still urge for explanations, especially in the neutrino sector. One of them is the smallness of neutrino masses, which are six orders of magnitude smaller than the electron mass. One elegant solution is to extend the SM with three right-handed (RH) neutrinos $\nu_R$, which are singlets under the SM gauge group. If the RH neutrinos acquire Majorana masses from a scale much higher than the electroweak scale, then they can naturally explain the observed feature of neutrino masses through the type-I seesaw mechanism \cite{Minkowski:1977sc,Gell-Mann:1979vob,Yanagida:1979as,Glashow:1979nm,Mohapatra:1979ia}.


Besides the smallness of neutrino masses, as more measurements in the neutrino sector have been conducted in the past few decades, two additional neutrino puzzles arise \cite{Mohapatra:2006gs,Xing:2020ijf}. One is the large mixing angles, which behave in a very different way compared to what we observed in the quark sector. The other is the weak hierarchy in the neutrino spectrum. Although the neutrino masses have not been measured directly, we can already tell that the two heavy neutrinos have their mass ratio $m_3/m_2 < 6$ regardless of the ordering, which is the smallest among all the SM species.


In this study, we try to answer two questions to figure out these neutrino puzzles. First, \textit{what is the origin of the right-handed neutrino masses and the seesaw scale?} The naive estimation of the seesaw scale $M_{\text{seesaw}}$ is $\lesssim 10^{14}$ GeV, which is much smaller than the Planck scale $M_{\text{Planck}}\sim 10^{19}$ GeV. The smallness of the ratio $M_{\text{seesaw}}/M_{\text{Planck}}$ is important to explain the observed neutrino masses, otherwise they will be even lighter. To generate such a ratio, one can refer to another successful story, Quantum Chromodynamics (QCD), which provides a remarkable and compelling way to generate the hierarchy between different scales. In QCD, the scale $\Lambda_{\text{QCD}}$ is introduced once the gauge coupling $g_s$ becomes strong through the logarithmic renormalization group running, which allows it to be far below the Planck scale $M_{\text{Planck}}$. Analogous to the $\Lambda_{\text{QCD}}$, the seesaw scale $M_{\text{seesaw}}$ could have a similar dynamical origin with an introduction of a new strong interaction. To construct a minimal model, a straightforward setup is to have RH neutrinos undergo the new strong interaction and acquire their Majorana masses dynamically through it.


Through the analogy to QCD, we get a convincing answer to the first question, a new strong interaction. However, there comes the second question. \textit{Can it also partly explain the large mixing and weak hierarchy observed from low-energy neutrino data?} To generate a nontrivial flavor structure of light neutrinos, the new strong interaction needs to know the flavor. A promising candidate is thus a horizontal gauge interaction \cite{Maehara:1978ts, Wilczek:1978xi} which becomes strongly coupled at the seesaw scale.\footnote{Strong horizontal gauge symmetries have also been studied for other purposes \cite{Hill:1990ge, King:1992qb, Elliott:1992xg, Elliott:1992ut}. An application on the seesaw model has been discussed in \cite{Smetana:2011tj, Hosek:2017ouz} but with different focuses and analyses from this study.} It is also the most economical choice because, in the extended SM with additional three generations of RH neutrinos, different generations are the only degrees of freedom for the RH neutrinos. The idea of relating the RH neutrino masses with horizontal gauge symmetry has already been mentioned in one of the first seesaw papers by Yanagida \cite{Yanagida:1979as}, where he used a scalar field to both generate the RH neutrino masses and break the horizontal gauge symmetry. It was already mentioned in the paper that the scalar can be a $\nu_R\overline{\nu_R^c}$ bound state, which means that the horizontal gauge symmetry is broken by the RH neutrino condensate through strong dynamics and the RH neutrino masses are dynamically generated, exactly what we propose. Our study can then be treated as a realization of this old idea with the new knowledge of the neutrino flavor structure from experiments.


Notice that we use the term ``\textit{partly}'' in the second question because, in the type-I seesaw mechanism, the resulting light neutrino masses and mixing will also depend on the Dirac neutrino masses originating from the Higgs field. These masses are determined by the Yukawa couplings between the Higgs field and the leptons, whose values are yet unknown. However, we do gain some hints about the pattern of Yukawa couplings from the quark sector, which suggests a hierarchical Dirac mass matrix with small mixing angles. Therefore, in this study, we assume that the Dirac mass matrix of neutrinos has a similar structure to that of up-type quarks, with a strong hierarchy and small mixing in the corresponding flavor basis.

Based on the assumption above, we discuss the RH neutrino mass matrix that could lead to the large mixing and weak hierarchy observed in the low-energy neutrino data. The idea is similar to the seesaw enhancement \cite{Smirnov:1993af}. It was shown that the RH neutrino mass matrix should have a special structure, either strongly hierarchical or strongly off-diagonal, to fit the data. To our interest, a strongly off-diagonal matrix is favored. We then examine the flavor structure of the RH neutrino condensate resulting from different choices of strong horizontal gauge symmetry. We find that a strongly anti-diagonal matrix, a special case of a strongly off-diagonal matrix, which emerges from $SU(2)_H$ gauge symmetry with RH neutrinos transforming as a triplet, can reproduce a light neutrino mass matrix consistent with the low-energy neutrino data.

From the analysis, we find that the flavor structure of neutrinos originates from some substructures inside the Dirac mass matrix beyond our assumptions which are enhanced by the strongly anti-diagonal flavor structure of the RH neutrino mass matrix. Therefore, the large mixing and weak hierarchy are the consequence of the substructures in the Dirac mass matrix $m_D$ with the strongly anti-diagonal RH mass matrix $M_R$ from the strong dynamics. In the end, most of the neutrino puzzles could be traced back to the RH neutrino condensate caused by the strong horizontal interaction at the seesaw scale.


This paper is organized as follows. We start from the formulation of our framework in section \ref{sec:Basic}. In section~\ref{sec:2G}, we first discuss a pedagogical case with only two generations of RH neutrinos to show the important features of the idea, including the strongly off-diagonal flavor structure and how it could originate from strong horizontal gauge symmetry. Next, a realistic model with three generations is presented in section~\ref{sec:3G}, where we find the scenario with the RH neutrinos in the $\bf{3}$ of $SU(2)_H$ horizontal gauge symmetry is favored. The realization of the seesaw enhancement through a strongly anti-diagonal matrix is discussed in detail with a benchmark model shown at the end. Moreover, we go up to construct a potential UV completion in section \ref{sec:UV} and go down to discuss relevant phenomenology that can be tested in section \ref{sec:Test}. Section~\ref{sec:Conclusion} contains our conclusions.

\section{The framework}\label{sec:Basic}

\subsection{Type-I seesaw mechanism with three generations of neutrinos}

In the type-I seesaw mechanism, we have a usual Dirac mass term for neutrinos as well as a Majorana mass term for the RH neutrinos given by the Lagrangian
\begin{align}
-\mathcal{L}_{m}&=\overline{\nu_L}\,m_D\,\nu_R+\frac{1}{2}\,\overline{\nu_R^c}\,M_R\,\nu_R+\text{h.c.} \nonumber\\
&=\frac{1}{2}
\begin{pmatrix}
\overline{\nu_{L}}  & \overline{\nu_{R}^c} 
\end{pmatrix}
\begin{pmatrix}
0     &  m_{D}  \\
m_{D}^T     &  M_R  \\
\end{pmatrix}
\begin{pmatrix}
\nu_{L}^c \\ \nu_{R} 
\end{pmatrix}
+\text{h.c.}~.
\label{Lseesaw0}
\end{align}
Since there are three generations of $\nu_L=(\nu_{L,1},\nu_{L,2},\nu_{L,3})$ and $\nu_R=(\nu_{R,1},\nu_{R,2},\nu_{R,3})$, both $m_D$ and $M_R$ are $3\times3$ mass matrices. The resulting light neutrino mass matrix $m_\nu$ from the type-I seesaw mechanism is then given by a $3\times3$ matrix satisfying
\begin{align}
m_\nu=-m_D\,{M_R^{-1}}\,{m_D^T}~.
\label{Mseesaw}
\end{align}

For the purpose of this study, we will work in the original flavor basis for all the fermions. That is, for example, the RH neutrinos are in the flavor eigenstate of horizontal gauge symmetry. In this basis, since the dynamics for the RH neutrino mass generation is known, the Majorana mass matrix $M_R$ will be in a simple form determined only by one mass parameter $M$.

 On the other hand, the dynamics for the Dirac mass matrix $m_D$ is still unknown so we can only write it in a general form with entries labelled by $m_{ij}$ where $i,j=1,2,3$. The resulting light neutrino mass matrix $m_\nu$ is labelled by $m_{ff'}$ where $f,f'=e,\mu ,\tau$. They can be expressed in matrix form as
 \begin{align}
 {m_D} =
\begin{pmatrix}
m_{11}   &  m_{12}  &  m_{13}   \\
m_{21}   &  m_{22}  &  m_{23}   \\
m_{31}   &  m_{32}  &  m_{33}   \\
\end{pmatrix}~,\quad
{m_\nu} =
\begin{pmatrix}
m_{ee}   &  m_{e\mu}  &  m_{e\tau}   \\
m_{\mu e}   &  m_{\mu\mu}  &  m_{\mu\tau}   \\
m_{\tau e}   &  m_{\tau\mu}  &  m_{\tau\tau}   \\
\end{pmatrix}~.
\label{matrix}
\end{align}
The Majorana mass matrix $M_R$ and $m_\nu$ should be symmetric with $m_{ff'}=m_{f'f}$ but the Dirac mass matrix $m_D$ does not need to.

\subsection{Dirac neutrino mass matrix analogous to up-type quark mass matrix}

Based on the current knowledge, there is no information on the exact structure of $m_D$. Therefore, some ad hoc assumptions are required to do a further analysis. In this study, we assume that in the flavor basis, the $m_D$ is similar to the Dirac mass matrix of up-type quarks. Such a relation can originate from quark-lepton symmetry or grand unified theories. However, we do not require the exact coincidence of the quark and lepton masses but just assume that there is a strong hierarchy of the eigenvalues as well as small mixing angles. We also borrow the notation from up-type quarks such that the eigenvalues of $m_D$ are given by $\{m_u,m_c,m_t\}$ or in matrix form as
\begin{align}
 {m_D^{\text{diag}}} =
\begin{pmatrix}
m_{u}   &  0  &  0   \\
0  &  m_{c}  &  0  \\
0   &  0 &  m_{t}   \\
\end{pmatrix}~.
\label{matrixmD}
\end{align}
For numerical study, we consider the up-type quark masses at the high scale $\sim 10^9$ GeV for the input \cite{Huang:2020hdv}, which can remove the QCD effect and reflect a more natural relation among masses of different generations in the lepton sector. The three Dirac neutrino masses are then given by
\begin{align}
m_{u}\sim 0.60 ~\text{MeV},\quad m_{c}\sim 400 ~\text{MeV},\quad m_{t}\sim 100 ~\text{GeV}~.
\end{align}

With these eigenvalues, we can try to reconstruct the Dirac neutrino mass matrix through the relation
\begin{align}
m_D = U_{L}~ {m_D^{\text{diag}}}~ U_{R}^\dagger ~,
\end{align}
where $U_L$ and $U_R$ are two different unitary matrices. However, the two matrices are still unknown even in the quark sector and the only information is that the $U_L$ of the up-type quark is related to the CKM matrix, which is parameterized by
\begin{align}
 U_{CKM} =
\begin{pmatrix}
c_{12}^Dc_{23}^D  &  s_{12}^Dc_{13}^D  &  s_{13}^De^{-i\delta_D}   \\
-s_{12}^Dc_{23}^D-c_{12}^Ds_{13}^Ds_{23}^De^{i\delta_D}  &  c_{12}^Dc_{23}^D-s_{12}^Ds_{13}^Ds_{23}^De^{i\delta_D}  &  c_{13}^Ds_{23}^D  \\
s_{12}^Ds_{23}^D-c_{12}^Ds_{13}^Dc_{23}^De^{i\delta_D}   &  -c_{12}^Ds_{23}^D-s_{12}^Ds_{13}^Dc_{23}^De^{i\delta_D} &  c_{13}^Dc_{23}^D   \\
\end{pmatrix}~,
\label{CKM}
\end{align}
where $c_{ij}^D\equiv\text{cos}\,\theta_{ij}^D$, $s_{ij}^D\equiv\text{sin}\,\theta_{ij}^D$, and $\delta_D$ is the CP-violating phase. The parameters are precisely measured through different experiments with current values given by \cite{ParticleDataGroup:2022pth}
\begin{align}
\theta^D_{12}=13.00^{\circ}~,\quad \theta^D_{23}=2.40^{\circ}~,\quad \theta^D_{13}=0.211^{\circ}~,\quad \delta_{D}={65.55}^{\circ}~,
\end{align}
which reflect a structure with small mixing angles.

In this work, to simplify the analysis, we assume the Dirac neutrino mass matrix $m_D$ is real and symmetric such that $m_D$ can be diagonalized by an orthogonal matrix $U_D$ and we further assume the $U_D$ follow the similar pattern as the CKM matrix such that $U_D \sim U_{CKM}$. With these assumptions, we can reconstruct the Dirac mass matrix through
\begin{align}
m_D \text{ (real-symmetric) } = U_{D}~ {m_D^{\text{diag}}}~ U_{D}^T ~,
\end{align}
which allows us to get a reasonable structure of $m_D$.\footnote{Notice that there will not be a complex phase in $U_D$ under this assumption, which makes it unrealistic. However, we claim it is still reasonable as in this work we focus on the consequence on the mixing angles and we expect the complex phase will not dramatically change our results. One can include the phase by relieving the assumption on $m_D$ from real-symmetric to hermitian. We leave a dedicated analysis on the role of the CP-violating phase for future study.} We will refer to these assumptions when encountering the Dirac mass matrix in the following sections.

\subsection{Light neutrino mass matrix from low-energy neutrino data}

The light Majorana neutrino mass matrix, on the other hand, must be symmetric and can be written through a unitary matrix $U_\nu$ and the diagonal mass matrix $m_\nu^{\text{diag}}$ as
\begin{align}
m_\nu = U_\nu^*~ {m_\nu^{\text{diag}}}~ U_\nu^\dagger ~.
\label{diagonalization}
\end{align}
The eigenvalues of $m_\nu$ are labelled by $\{m_1,m_2,m_3\}$ and the relation between the eigenvalues can be extracted from the neutrino oscillation data. Taking the recent global fits for the neutrino oscillation parameters \cite{Esteban:2020cvm, Esteban:2022}, the two mass-squared differences are given by
\begin{align}
\Delta m_{12}^2\equiv m_2^2-m_1^2 = 74.1 \text{  meV}^2, ~\quad~
\Delta m_{3\ell}^2\equiv m_3^2-m_1^2 = 2511 \text{  meV}^2 ~,
\label{Deltam2}
\end{align}
where we consider the normal ordering without the Super-Kamiokande data. From these values, the weak hierarchy of the neutrino spectrum can be shown as
\begin{align}
m_3/m_2 ~\leq~ \sqrt{\Delta m_{3\ell}^2\,/\,\Delta m_{12}^2}~=~5.8~.
\end{align}

For the unitary matrix $U_\nu$, the neutrino oscillation experiments can also provide some information. The leptonic mixing matrix, also known as the Pontecorvo-Maki-Nakagawa-Sakata (PMNS) mixing matrix $U_{PMNS}$ \cite{Pontecorvo:1957cp, Pontecorvo:1957qd, Maki:1962mu}, is parameterized by
\begin{align}
 U_{PMNS} =
\begin{pmatrix}
c_{12}c_{23}   &  s_{12}c_{13}  &  s_{13}e^{-i\delta}   \\
-s_{12}c_{23}-c_{12}s_{13}s_{23}e^{i\delta}  &  c_{12}c_{23}-s_{12}s_{13}s_{23}e^{i\delta}  &  c_{13}s_{23}  \\
s_{12}s_{23}-c_{12}s_{13}c_{23}e^{i\delta}   &  -c_{12}s_{23}-s_{12}s_{13}c_{23}e^{i\delta} &  c_{13}c_{23}   \\
\end{pmatrix}\cdot
\begin{pmatrix}
e^{i\eta_1}   &  0  &  0  \\
0   &  e^{i\eta_2}  &  0   \\
0   &  0 &  1   \\
\end{pmatrix},
\label{PMNS}
\end{align}
where $c_{ij}\equiv\text{cos}\,\theta_{ij}$, $s_{ij}\equiv\text{sin}\,\theta_{ij}$, $\delta$ is the CP-violating phase, and $\eta_1$, $\eta_2$ are the Majorana phases of $m_1$ and $m_2$. The first three parameters can be measured through the neutrino oscillation experiments. Taking the recent global fits for the neutrino oscillation parameters \cite{Esteban:2020cvm, Esteban:2022}, the mixing angles and the CP-violating phase are given by
\begin{align}
\theta_{12}=33.41^{\circ}~,\quad \theta_{23}=49.1^{\circ}~,\quad \theta_{13}=8.54^{\circ}~,\quad \delta_{}={197^{+42}_{-25}}^{\,\circ}~,
\end{align}
where we show the uncertainty of the CP-violating phase due to its larger value (other uncertainties are within $\sim 5\%$). The two Majorana phases are unknown.

The PMNS matrix shows a totally different pattern compared to the CKM matrix, especially the two large mixing angles $\theta_{12}$ and $\theta_{23}$. However, it is not enough to get $U_\nu$ because the PMNS matrix is determined by both charged lepton rotation and neutrino rotation as $U_{PMNS}=U_\ell^\dagger U_\nu$, where $U_\ell$ is the unitary matrix satisfying the diagonalization of the charged lepton mass matrix $U_\ell^\dagger \,m_\ell\, V_\ell = m_\ell^{\text{diag}}$. Since we assume a hierarchical Dirac mass matrix with small mixing angles for neutrinos, we expect a similar structure for the charged lepton mass matrix $m_\ell$. Therefore, we also assume the matrix $U_\ell \sim U_{CKM}$. Under this assumption, the main contribution for the large-mixing structure of $U_{PMNS}$ should come from the neutrino mixing matrix as $U_{PMNS}=U_\ell^\dagger U_\nu \sim U_{CKM}^\dagger U_\nu \sim U_\nu$.

With information on both $m_\nu^{\text{diag}}$ and $U_\nu$, we can now derive the flavor structure of $m_\nu$. The goal of this study is to generate a $m_\nu$ consistent with this flavor structure from the strong interaction induced type-I seesaw mechanism.


\section{First attempt: start with two generations of right-handed neutrinos}\label{sec:2G}

To get the basic concept of seesaw enhancement and strong horizontal gauge interactions, we start from a simple attempt with only two generations, where we use the second and third generations labeled by $\nu_{L/R,2}$ and $\nu_{L/R,3}$. For the first attempt, We proceed with the bottom-up approach in this section, where we first discuss the desired mass matrix for $M_R$ and then examine the possible candidates for the strong interaction.

\subsection{Large mixing and weak hierarchy from a strongly off-diagonal structure}\label{sec:IR2}

To begin with, we show how a strongly off-diagonal structure of $M_R$ can explain the neutrino flavor puzzles through the seesaw enhancement \cite{Smirnov:1993af}. Assume the RH neutrinos get masses at the seesaw scale with a value of $M$ only in the off-diagonal entries of $M_R$. The $4\times 4$ mass matrix for the neutrinos is then given by
\begin{align}
-\mathcal{L}_{m,2G}=\frac{1}{2}
\begin{pmatrix}
\overline{\nu_{L,2}} & \overline{\nu_{L,3}} & \overline{\nu_{R,2}^c} & \overline{\nu_{R,3}^c}
\end{pmatrix}
\begin{pmatrix}
0   &  0  &  m_{22}   &  m_{23}  \\
0   &  0  &  m_{32}   &  m_{33}  \\
m_{22}   &  m_{32}  &  0   &  M  \\
m_{23}  &  m_{33}  &  M   &  0  \\
\end{pmatrix}
\begin{pmatrix}
\nu_{L,2}^c \\ \nu_{L,3}^c \\ \nu_{R,2} \\ \nu_{R,3}
\end{pmatrix}
+\text{h.c.}~.
\label{Lmass2}
\end{align}
For the Dirac mass matrix, we assume it is real and symmetric, $m_{23}=m_{32}$, and follows the same structure as the up-type quark. That is, the eigenvalues of the $2\times2$ matrix should be $m_c\sim m_{22}$ and $m_t \sim m_{33}$ with only a small rotation sin$\,\theta_{23}^D \lesssim 0.04$ analogous to the CKM matrix. Since the exact mixing angle is unknown, we discuss three different cases :
\\
\\
(1) Starting with the case where the Dirac mass matrix is exactly diagonal with $m_{23}=0$ and thus $m_{22}=m_c$, $m_{33}=m_t$, we can derive the light neutrino mass matrix as
\begin{align}
{m_\nu}\sim -\frac{1}{M}
\begin{pmatrix}
0   &  m_{22}m_{33}    \\
m_{22}m_{33}   &  0    \\ 
\end{pmatrix}
\implies
m_2=m_3=\frac{m_{22}m_{33}}{M}=\frac{m_{c}m_{t}}{M}\ ,~
\text{sin}^2\,\theta_{23} = \frac{1}{2}~.
\end{align}
We end up with two degenerate light neutrinos with maximal mixing. Also, the Majorana phase is maximal, which means the two neutrinos have opposite CP parities.
\\
\\
(2) If the off-diagonal term $m_{23}$ is nonzero but relatively small ($m_{23} \ll m_{22}\sim m_c$), the spectrum will split with the new mass matrix and eigenvalues given by
\begin{align}
{m_\nu}\sim -\frac{1}{M}
\begin{pmatrix}
0   &  m_{22}m_{33}    \\
m_{22}m_{33}   &  2m_{23}m_{33}    \\
\end{pmatrix}
\implies
m_2\sim \frac{m_{33}}{M}\left(m_{22}-m_{23}\right), ~
m_3\sim \frac{m_{33}}{M}\left(m_{22}+m_{23}\right)\ .
\end{align}
The angle will also shift away from the maximum. The split in the mass spectrum and the deviation from the maximal mixing can be written as
\begin{align}
R_{32} \equiv \frac{m_3-m_2}{m_3+m_2}\sim\frac{m_{23}}{m_{22}}\ ,\quad
\text{sin}^2\,\theta_{23} \sim \frac{1}{2}\left(1-\frac{m_{23}}{m_{22}}\right)\ , 
 \end{align}
where both values are related to the small ratio $m_{23}/m_{22}$. Therefore, if this is the case, we should find a quasi-degenerate spectrum with the splitting directly related to the angle.
\\
\\
(3) The value of $m_{23}$ can be as large as $m_{23}=m_t\,$sin$\,\theta_{23}^D \sim 10\, m_c$ within our assumptions. For a $m_{23}$ comparable to $m_{22}$, the term $2m_{23}m_{33}$ at the diagonal entry will become relevant, which can generate a rather hierarchical neutrino spectrum with a small mixing. We can define the ratio $N\equiv m_{23}/m_{22}$. The spectrum and the mixing are given by
\begin{align}
m_2 \sim \frac{m_{22}m_{33}}{M}\left(\sqrt{N^2+1}-N\right), \quad 
m_3 \sim \frac{m_{22}m_{33}}{M}\left(\sqrt{N^2+1}+N\right)\ ,\quad
\text{tan}\,2\theta_{23} = \frac{1}{N}~.
\end{align}
The weak hierarchy condition $m_3/m_2 < 6$ thus requires $N\equiv m_{23}/m_{22} < 1$.
\\

Besides a strongly off-diagonal structure, the seesaw enhancement can also be realized through a diagonal RH neutrino mass matrix $M_R$ with strongly (quadratically) hierarchical entries \cite{Smirnov:1993af}. The essence is to compensate the hierarchy in the Dirac mass matrix $m_D$ to get a quasi-democratic structure of the mass matrix $m_\nu$. However, in the next subsection, we will show why it is not favored in a strongly coupled theory. The other problem of the seesaw enhancement through a strongly hierarchical structure is that fine-tuning is required to get perfect compensation. The problem is inevitable in the generic type-I seesaw model. It can be explained in the double seesaw model \cite{Lindner:2005pk,Kim:2005kca}, where the RH neutrinos also get their masses from the seesaw mechanism, but it will require a more complicated setup.

\subsection{Flavor structures from strong horizontal gauge symmetry with two $\nu_R$'s}\label{sec:UV2}

To generate the RH neutrino mass matrix with a nontrivial flavor structure dynamically, a flavorful strong interaction is required. In this subsection, we examine all the possible strong horizontal gauge interactions to assess their capability to fulfill this task. For the pedagogical case with only two generations of RH neutrinos, there are several candidates for horizontal gauge symmetry, including $SO(2)_H$, $SU(2)_{H}$, and $U(1)_{H}$.\footnote{The $U(1)_{H}$ and $SO(2)_H$ are isomorphic. However, the fermion doublet under $SO(2)_H$ and $U(1)_{H}$ (the pair of fermions carrying charge $\pm 1/2$) are different up to a rotation of basis. Such a rotation changes the flavor basis, which has a dramatic impact on the analysis of flavor structure. Therefore, although they are isomorphic, we distinguish them as two different scenarios.}

Start with $SO(2)_{H}$ horizontal gauge symmetry where $\nu_{R,2}$ and $\nu_{R,3}$ transform as a fundamental representation $\bf{2}$. The interaction term is given by 
\begin{equation}
\mathcal{L}_{\text{SO(2)}}=\frac{i}{2}\,g_H Z_H^\mu \left(\overline{\nu_{R,3}}\gamma_\mu {\nu_{R,2}}-\overline{\nu_{R,2}}\gamma_\mu {\nu_{R,3}}\right)~,
\end{equation}
where $g_H$ is the gauge coupling and $Z_H$ is the horizontal gauge boson. Once $g_H$ becomes strong, the RH neutrinos will condense. To check the properties of the condensate, we need to know which combination of fermion bilinears $\nu_{R}\overline{\nu_{R}^c}$ gives the true vacuum. In this study, we follow the criterion of the most attractive channel (MAC) \cite{Raby:1979my}. With the given representations $R_1$, $R_2$, and $R_3$, the relative strength of certain condensate can be derived as $V=C_1+C_2-C_3$, where $C_i$ is the quadratic Casimir invariant of the representation $R_i$. For the combination of two $\bf{2}$'s, we have two possible fermion bilinears as $\bf{2}\times \bf{2}\to \bf{3}+\bf{1}$. The criterion says the most attractive channel should be $\bf{2}\times \bf{2}\to \bf{1}$. Conversely, $\bf{2}\times \bf{2}\to \bf{3}$ with a negative strength is actually a repulsive channel.

After determining the flavor structure, we also need to check the spin and spatial parts. Since we have a condensate formed by two identical fermions, the overall wavefunction for the RH neutrino bilinear should be antisymmetric due to the Pauli exclusion principle. The flavor wavefunction of the singlet $\bf{1}$ from $\bf{2}\times \bf{2}$ under $SO(2)_H$ is symmetric. Therefore, the combined spin and spatial wavefunction must be antisymmetric with the lowest potential, which will be the Lorentz scalar condensate in this case.

The resulting scalar condensate can be described by an auxiliary scalar field $\Phi =  \nu_{R}\overline{\nu_{R}^c}$ with a nonzero vacuum expectation value (VEV) $\langle\Phi\rangle =f$. The relevant Lagrangian for the dynamical mass generation is given by
\begin{align}
-\mathcal{L}_{\Phi}=\frac{1}{2}y_s
\begin{pmatrix}
\overline{\nu_{R,2}^c} & \overline{\nu_{R,3}^c}
\end{pmatrix}
\Phi_{2\times2}
\begin{pmatrix}
\nu_{R,2} \\ \nu_{R,3}
\end{pmatrix}
+\text{h.c.}~ \quad \implies \quad M_R=y_s \langle \Phi_{2\times2}\rangle ~,
\label{scalar}
\end{align}
where $y_s$ is the Yukawa coupling between the bound state and its component with an $O(1)$ value determined by the strong dynamics. The Majorana masses are generated through the scalar's VEV with $M= y_sf$. Then, since the singlet $\bf{1}$ from $\bf{2}\times \bf{2}$ under $SO(2)_H$ has their flavor index contracted via a delta function, the flavor structure of the mass matrix $M_R$ can be expressed in the flavor basis given by
\begin{align}
\nu_R = \mathbf{2} \text{ of } SO(2)_H ~:
\quad 
\langle \Phi_{2\times2}\rangle=
\begin{pmatrix}
  f   &  0  \\
  0   &  f  \\
\end{pmatrix}
\quad \implies \quad M_R=
\begin{pmatrix}
  M   &  0  \\
  0   &  M  \\
\end{pmatrix}~.
\label{MRSU2}
\end{align}
The mass matrix $M_R$ shares the same flavor structure as $\langle\Phi\rangle$ with only diagonal entries, which, however, is contrary to the desired strongly off-diagonal structure.

The next candidate is $SU(2)_H$ horizontal gauge symmetry with RH neutrinos transforming as a doublet $\bf{2}$. The interaction term is given by 
\begin{equation}
\mathcal{L}_{\text{SU(2)}}=\frac{1}{2}\,g_H W_H^{a,\,\mu} \left(\overline{\nu_{R}}\,\gamma_\mu \sigma^a\, {\nu_{R}}\right)~,
\end{equation}
where $W_H^a$ are the $SU(2)_H$ gauge bosons and $\sigma^a$ are Pauli matrices with $a=1,2,3$. Similar to the previous case, once $g_H$ becomes strong, it will lead to a condensate between $\nu_R$'s. The most attractive channel for doublets $\bf{2}$'s of $SU(2)_H$ is also $\bf{2}\times \bf{2}\to \bf{1}$. However, there is a crucial distinction between the two gauge groups. That is, a $\bf{2}$ of $SO(2)$ is a real representation, whereas a $\bf{2}$ of $SU(2)$ is a pseudo-real representation, which leads to a huge difference in the flavor structure. Because of the pseudo-real nature, the flavor index is contracted via an antisymmetric epsilon, which will give a condensate with only off-diagonal entries in the flavor basis as desired. However, due to the Pauli exclusion principle, the combined spin and spatial wavefunction of the condensate now needs to be symmetric and therefore will not be a Lorentz scalar. Such a condensate will thus break the Lorentz symmetry which is unacceptable.

It looks like we are only a minus away from the desired flavor structure, which can be achieved by keeping only the $a=3$ component of $SU(2)_H$ interactions. The interaction is an $U(1)_{H}$ gauge symmetry where $\nu_{R,2}$ carries charge $1/2$ and $\nu_{R,3}$ carries charge $-1/2$, similar to $U(1)_{L_\mu-L_\tau}$, which is given by 
\begin{equation}
\mathcal{L}_{\text{U(1)}}=\frac{1}{2}g_H W_H^{3,\,\mu} \left(\overline{\nu_{R,2}}\gamma_\mu {\nu_{R,2}}-\overline{\nu_{R,3}}\gamma_\mu {\nu_{R,3}}\right)~.
\end{equation}
The condensate can then be a Lorentz scalar and symmetric under the flavor basis. Again, we can use an auxiliary scalar field $\Phi =  \nu_{R}\overline{\nu_{R}^c}$ to describe it. The flavor structure of the VEV is now given by
\begin{align}
\nu_R = \left(\frac{1}{2},-\frac{1}{2}\right) \text{ of } U(1)_H ~:
\quad 
\langle \Phi_{2\times2}\rangle=
\begin{pmatrix}
  0   &  f  \\
  f   &  0  \\
\end{pmatrix}
\quad \implies \quad M_R=
\begin{pmatrix}
  0   &  M  \\
  M   &  0  \\
\end{pmatrix}~,
\label{MRU1}
\end{align}
which then gives us a strongly off-diagonal structure for $M_R$ as we want. Under the strong $U(1)_H$ interaction, such a strongly off-diagonal feature is simply the consequence of charge assignments. The condensates only happen when combining neutrinos with opposite charges, which are $\nu_{R,2}\overline{\nu_{R,3}^c}$ and $\nu_{R,3}\overline{\nu_{R,2}^c}$ in this case. On the other hand, the diagonal entries are zero because they are repulsive channels. If we look into the resulting mass term in eq.~\eqref{MRU1}, it is actually a Dirac mass among $\nu_{R_2}$ and $\nu_{R_3}$. It makes sense as the RH neutrinos carry charge under $U(1)_H$ so Majorana masses are forbidden and the masses we get must be Dirac masses. Such a nature is also reflected on the degenerate spectrum and the maximal mixing for the RH neutrinos.

One common feature of dynamical mass generation from these strongly coupled theories is that, regardless of the flavor structure, they always lead to a condensate with two degenerate RH neutrinos. Such a feature originates from the stability of the vacuum from the strong dynamics. The resulting RH neutrino mass matrix is characterized by a single mass scale $M$. Therefore, in the two-generation scenario, a strongly off-diagonal structure, which can achieve the seesaw enhancement with one mass parameter, is favored over a strong hierarchical diagonal structure, which requires additional mass parameters.

The other common feature is the global symmetry of the theory, which is RH neutrino number $U(1)_R$ in all the scenarios. The $U(1)_R$ symmetry is broken by the RH neutrino condensate. This kind of breaking is not presented in the SM because QCD is a vector theory with its vacuum preserving the baryon/quark number. In our study, all the scenarios above are anomaly-free chiral gauge theories, which provide a chance to form a RH neutrino number violating condensate. Such a chiral feature is the key to a successful dynamical generation of Majorana masses. The breaking of the global $U(1)_R$ symmetry also implies the existence of a massless Goldstone boson, known as majoron \cite{Chikashige:1980ui}, which is a bound state of RH neutrinos in our models.

To sum up, we find that the first two horizontal gauge symmetry candidates, $SO(2)_H$ and $SU(2)_H$, will lead to either a wrong flavor structure or a wrong Lorentz structure. It turns out that the most promising choice is a strong $U(1)_H$ horizontal gauge symmetry, which can generate the desired mass matrix $M_R$ in the scenario with two generations of RH neutrinos. However, using the $U(1)_H$ gauge symmetry as a strong interaction is not an ideal choice because it is not asymptotically free and further efforts are required to explain how the gauge coupling becomes strong in the infrared.\footnote{A strong $U(1)$ gauge interaction has also been used to generate the dynamical electroweak symmetry breaking in class of top quark condensation models \cite{Lindner:1991bs,Giudice:1991sz}. However, none of them explains how to get a strongly coupled $U(1)$ gauge group in the infrared.} Therefore, we will not consider $U(1)_H$ gauge interactions in the following study.

The failure in the first attempt with only two generations of RH neutrinos shows how restricted the model is. The restriction arises primarily because strong dynamics fix the structure of the condensate, resulting in a loss of freedom to tune the structures and parameters. However, in the next section, we will see that there is an additional candidate in the three-generation scenario, which can generate a promising RH neutrino mass matrix and lead to a light neutrino mass matrix consistent with the data.


\section{A realistic model with three generations of right-handed neutrinos}\label{sec:3G}

In this section, to construct a realistic model, we adopt a top-down approach by first identifying the viable strong horizontal gauge symmetry. Subsequently, based on the resulting RH neutrino mass matrix $M_R$, we delve into the seesaw enhancement and construct the desired Dirac mass matrix $m_D$. The benchmark model derived from a numerical study will be presented in the end.

\subsection{Flavor structures from strong horizontal gauge symmetry with three $\nu_R$'s}\label{sec:UV3}

Most of the statements in the previous section can be directly extrapolated to the three-generation scenario. The horizontal gauge symmetry $SO(3)_H$ with $\nu_{R}=\bf{3}$ will still generate a singlet condensate via $\bf{3}\times \bf{3} \to \bf{1}$, resulting in a diagonal RH neutrino mass matrix $M_R$. This configuration, however, fails to generate the large mixing and weak hierarchy with the assumption of a hierarchical Dirac mass matrix.

The $SU(3)_H$ with $\nu_{R}=\bf{3}$ has RH neutrinos now in a complex representation. However, the outcome remains similar. The $\bf{3}\times \bf{3}$ combination under $SU(3)_H$ can lead to either a symmetric $\bf{6}$ or an anti-symmetric $\bf{\bar{3}}$. Following the criterion of the most attractive channel, we find the condensate should form in the $\bf{3}\times \bf{3}\to \bf{\bar{3}}$ channel. However, the anti-symmetric flavor wavefunction will again force the condensate to break the Lorentz symmetry, which is not allowed. Notice that the $\bf{3}\times \bf{3}\to \bf{6}$ with a negative strength is actually repulsive. Therefore, the statement in the original seesaw paper by Yanagida \cite{Yanagida:1979as} with a RH neutrino bound state $\chi=\nu_{R}\overline{\nu_{R}^c}=\bf{6}$ of $SU(3)_H$ can not be realized in our setup.

However, there is an additional choice for the three-generation scenario. That is having three RH neutrinos transforming as a triplet $\bf{3}$ of $SU(2)_H$.\footnote{The $SU(2)_H$ horizontal gauge symmetry has already been used in model building to understand the flavor puzzles since a long time ago \cite{Babu:1990hu,Shaw:1992gk}, including some studies focusing on the lepton mixing \cite{Kuchimanchi:2002fi,Kuchimanchi:2002yu}. The idea is also revisited in some recent works \cite{Darme:2023nsy,Greljo:2023bix}. However, most of them are based on the assumption that the three generations of fermions transform as a $\bf{2}+\bf{1}$ instead of $\bf{3}$.} Among the different combinations $\bf{3}\times \bf{3}\to \bf{5}+\bf{3}+\bf{1}$, the most attractive channel is $\bf{3}\times \bf{3}\to \bf{1}$. Since the $\bf{3}$ is in a real representation, the singlet condensate will be symmetric in the flavor basis. Therefore, a Majorana-type scalar condensate is allowed, different from the cases with $\bf{2}$ of $SU(2)_H$ and $\bf{3}$ of $SU(3)_H$. Compared to $U(1)_H$, the non-abelian $SU(2)_H$ can become strongly coupled in the infrared if the particles charged under the $SU(2)_H$ are not too numerous.

To understand the flavor structure, we can use the auxiliary scalar field $\Phi =  \nu_{R}\overline{\nu_{R}^c}$ again. The scalar gets a VEV which is a singlet $\bf{1}$ of $SU(2)_H$. We can then express the VEV and the resulting RH mass matrix in the flavor basis as
\begin{align}
\nu_R = \mathbf{3} \text{ of } SU(2)_H ~:\quad 
\langle \Phi_{3\times3}\rangle=
\begin{pmatrix}
  0   &  0   &  f  \\
  0   &  -f  &  0  \\
  f    &  0   & 0  \\
\end{pmatrix}\quad\implies\quad M_R=
\begin{pmatrix}
  0   &  0   &  M  \\
  0   &  -M  &  0  \\
  M    &  0   & 0  \\
\end{pmatrix}
\label{MR3}~,
\end{align}
where the mass $M=y_sf$ with $y_s$ determined by the $SU(2)_H$ strong dynamics. We get the RH neutrino mass matrix $M_R$ with a promising strongly anti-diagonal structure, which is similar to the strongly off-diagonal structure. We then want to check if we can successfully reproduce the desired light neutrino flavor structure with this $M_R$.

\subsection{Seesaw enhancement through a strongly anti-diagonal mass matrix}\label{sec:SSE}

The strongly anti-diagonal mass matrix $M_R$ provides a promising flavor structure to realize a successful seesaw enhancement. A comprehensive study of seesaw enhancement with three RH neutrinos has already been done in \cite{Akhmedov:2003dg}, where a bottom-up approach is used to probe the structure of $M_R$. Our model with three degenerate RH neutrinos can fit into the special case III in their study, where the main structure of the RH neutrino mass matrix is similar to the strongly anti-diagonal $M_R$ in eq.~\eqref{MR3}. However, in this work, we adopt a distinct assumption on the basis, leading to a different analysis.

In \cite{Akhmedov:2003dg}, the authors also assume the Dirac neutrino mass matrix $m_D$ is similar to the up-type quark mass matrix but work in the basis where $m_D$ is exactly diagonal. Therefore, the required RH mass matrix needs to have some subleading terms in order to reproduce the observed neutrino data. Moreover, the values of these subleading terms need to be strongly hierarchical to achieve this goal, which reintroduces a fine-tuning issue.

In this study, we have RH neutrino masses dynamically generated from the known strong dynamics so the mass matrix $M_R$ is merely determined by one parameter $M$ without any subleading structures. On the other side, we consider a non-diagonal Dirac neutrino mass matrix $m_D$ in the flavor basis. In the following analysis, we will show that these off-diagonal terms, though constrained by the small CKM-like mixing, can result in the large mixing and weak hierarchy in the neutrino sector.

Comparing the two analyses, the subleading terms in the RH neutrino mass matrix are trade-off by the off-diagonal terms in the Dirac mass matrix. Numerically, one can also interpret them as analyses conducted in two different flavor bases. However, the meaning is quite different from the top-down perspective. As we will demonstrate later, the seesaw enhancement now implies the enhancement of the flavor substructure inside the Dirac mass matrix $m_D$ through the strongly anti-diagonal RH neutrino mass matrix $M_R$. We also claim that it is closer to the natural flavor basis.

Based on our assumptions, we start with a non-diagonal, real-symmetric Dirac mass matrix $m_D$. The $6\times 6$ neutrino mass matrix with $M_R$ from $SU(2)_H$ strong dynamics given in eq.~\eqref{MR3} is
\begin{align}
-\mathcal{L}_{m,3G}=\frac{1}{2}
\begin{pmatrix}
\overline{\nu_{L,1}} & \overline{\nu_{L,2}} & \overline{\nu_{L,3}} & \overline{\nu_{R,1}^c} & \overline{\nu_{R,2}^c} & \overline{\nu_{R,3}^c}
\end{pmatrix}
\begin{pmatrix}
0   &  0  & 0  &  m_{11}   &  m_{12}  &  m_{13}\\
0   &  0  & 0  &  m_{12}   &  m_{22}  &  m_{23}\\
0   &  0  & 0  &  m_{13}   &  m_{23}   &  m_{33}  \\
m_{11}   &  m_{12}  &  m_{13}  &  0  &  0  &  M \\
m_{12}   &  m_{22}  &  m_{23}  &  0  & -M   &  0  \\
m_{13}   &  m_{23}  &  m_{33}  &  M  &  0  &  0  \\
\end{pmatrix}
\begin{pmatrix}
\nu_{L,1}^c \\\nu_{L,2}^c \\ \nu_{L,3}^c \\ \nu_{R,1} \\ \nu_{R,2} \\ \nu_{R,3}
\end{pmatrix}
+\text{h.c.}
\label{Lmass3}
\end{align}
The resulting light neutrino mass matrix from the type-I seesaw mechanism is given by
\begin{align}
{m_\nu}\sim \frac{1}{M}
\begin{pmatrix}
m_{12}^2-2m_{11}m_{13}~    &  m_{12}m_{22}-m_{11}m_{23}-m_{12}m_{13}    &  -m_{11}m_{33} + m_{12}m_{23} -m_{13}^2   \\
...  &  m_{22}^2-2m_{12}m_{23}  &  m_{22}m_{23} - m_{12}m_{33} - m_{13}m_{23}   \\
...   &  ...  &  m_{23}^2-2m_{13}m_{33}   \\
\end{pmatrix}.
\end{align}
One can easily check that if the Dirac mass matrix $m_D$ is diagonal, the resulting mass matrix $m_\nu$ will also be anti-diagonal with only three nonzero entries, which are $m_{\mu\mu}=m_{22}^2/M$ and $m_{e\tau}=m_{\tau e}=-m_{11}m_{33}/M$. In this case, the resulting $\nu_1$ and $\nu_3$ are degenerate and maximally mixed with $\nu_2$ decoupled, which is inconsistent with the data. Therefore, the off-diagonal terms in the Dirac mass matrix $m_D$ play an important role in helping get the correct neutrino flavor structure.

To discuss the impact of the off-diagonal terms, we follow the assumptions that the Dirac mass matrix $m_D$ is similar to the up-type quark sector with hierarchical eigenvalues and small mixing angles. The values of off-diagonal terms are thus restricted and can not be arbitrary large. The direct consequence of the assumption is that $m_{ee}$ and $m_{e\mu}$ are rather small, which matches the bottom-up prediction from \cite{Akhmedov:2003dg}. From a bottom-up point of view, the condition looks fine-tuned but it is actually a natural consequence in our model with the mass matrix $M_R$ given in eq.~\eqref{MR3} and a hierarchical Dirac mass matrix $m_D$.

There are two further phenomenological consequences of this feature in neutrino physics. First, combined with the existing neutrino data, the neutrino masses should be in normal ordering ($m_3>m_2>m_1$). Second, if the LH charged lepton rotation is zero, the value of $m_{ee}$ will directly match the mass $m_{\beta\beta}$ measured in the neutrinoless double beta decay experiments. Even with small LH charged lepton mixing analogous to the CKM matrix, $m_{\beta\beta}$ will still be negligible due to small $m_{ee}$ and $m_{e\mu}$. Therefore, if any large signal is found in this type of measurement, the idea will be strongly constrained. More phenomenology will be discussed in section \ref{sec:Test}.

\subsection{Analytical study and substructures inside the Dirac mass matrix}\label{sec:SSE}

Yet we get a light neutrino mass matrix $m_\nu$ in terms of the RH neutrino mass $M$ and the Dirac mass matrix elements. To generate the desired large mixing and weak hierarchy, further assumptions on the substructures of Dirac mass matrix are required.

As we already know the two entries $m_{ee}$ and $m_{e\mu}$ are small, the rest of the entries in $m_\nu$ should be of the same order. To achieve this, we start with the following relations among the Dirac mass matrix elements
\begin{align}
m_{33}\sim m_t~,\quad m_{22}\sim m_{23}\sim  m_c~, \quad m_{11}\sim m_{12}\sim m_{13}\sim m_u~.
\end{align}
These values can be parameterized by a small factor $\epsilon\sim O(0.01)$ as $m_c/m_t\sim m_u/m_c \sim \epsilon$ approximately. Then, the Dirac mass matrix can be expressed as
\begin{align}
{m_D}=
\begin{pmatrix}
m_{11}   &  m_{12}  &  m_{13}   \\
m_{12}   &  m_{22}  &  m_{23}   \\
m_{13}   &  m_{23}  &  m_{33}   \\
\end{pmatrix}\sim {m_t}
\begin{pmatrix}
\epsilon^2    &  \epsilon^2    &  \epsilon^2   \\
\epsilon^2  &  \epsilon  &  \epsilon   \\
\epsilon^2   &  \epsilon  &  1   \\
\end{pmatrix}~\implies~
{m_D^{\text{diag}}} \sim {m_t}
\begin{pmatrix}
\epsilon^2    &  0  &  0   \\
0  &  \epsilon   &  0  \\
0   &  0 &  1   \\
\end{pmatrix}\label{mD}
\end{align}
with three small mixing angles satisfying sin$\,\theta^D_{12}\sim$ sin$\,\theta^D_{23}\sim \epsilon$ and sin$\,\theta^D_{13}\sim \epsilon^2$. The resulting light neutrino mass matrix has an approximate form as
\begin{align}
{m_\nu}\sim \frac{m_t^2}{M}
\begin{pmatrix}
0    &  0    &  -\epsilon^2   \\
0  &  \epsilon^2  &  \epsilon^2   \\
-\epsilon^2   &  \epsilon^2  &  \epsilon^2   \\
\end{pmatrix}+O(\epsilon^3)~,
\label{mnu}
\end{align}
which again show the small $m_{ee}$ and $m_{e\mu}$. Besides, the matrix of the leading-order $\epsilon^2$ terms is the same as the $A_1$ mass matrix with two-zero textures in \cite{Frampton:2002yf}. To further analyze the flavor structure, we look into the leading term in each entry, which are given by
\begin{align}\label{mL}
{m_{\nu}}\text{ (leading term) }\sim \frac{1}{M}
\begin{pmatrix}
0    &  0    &  -m_{11}m_{33}   \\
0    &  m_{22}^2  &  m_{22}m_{23}   \\
-m_{11}m_{33}~   &  m_{22}m_{23}  &  m_{23}^2   \\
\end{pmatrix}~,
\end{align}
where we consider the ordering $m_{22}m_{23}>m_{12}m_{33}$ and $m_{23}^2>m_{13}m_{33}$ among the $O(\epsilon^2)$ terms in this analysis.

To reproduce the correct spectrum and mixing, we need the 13 entry to be small due to the observation of small $\theta_{13}$. Under this assumption, the bottom-right $2\times 2$ block becomes dominant, which gives
\begin{align}
\text{tan}\,\theta_{23} = \frac{m_{22}}{m_{23}}\ ,\quad
m_3\sim\frac{1}{M}(m_{22}^2+m_{23}^2)~.
\end{align}
The large $\theta_{23}$ mixing thus implies $m_{22} \sim m_{23} \sim m_c$, which serves as the first relation between the neutrino mixing and the substructure inside the Dirac mass matrix. That is, \textit{the large mixing $\theta_{23}$ comes from the nearly degenerate $m_{22} \sim m_{23}$ in the Dirac neutrino mass matrix.} Also, the relation allows us to settle down the mass scale of the RH neutrinos (the seesaw scale) as  
\begin{align}
M \sim \frac{m_{22}^2+m_{23}^2}{m_3}=6.4\times 10^9 \text{ GeV }
\left(\frac{m_c}{400\text{ MeV}}\right)^2\left(\frac{50\text{ meV}}{m_3}\right)~.
\end{align}

Next, we can also estimate the other two mixing angles, which are given as
\begin{align}
\text{sin}\,\theta_{12} \sim \frac{1}{\sqrt{2}}\ ,\quad
\text{sin}\,\theta_{13} \sim \frac{m_{11}m_{33}}{m_{22}^2+m_{23}^2}\text{ cos}\,\theta_{23} = \frac{m_{11}m_{33}}{m_{22}^2+m_{23}^2}\frac{m_{23}}{\sqrt{m_{22}^2+m_{23}^2}} ~.
\end{align}
We get another large mixing angle $\theta_{12}$ which is already embedded in the matrix \eqref{mL} with a dominant bottom-right $2\times 2$ block. The other angle, on the other hand, depends on the mass parameter in the Dirac mass matrix. As we already know $m_{22} \sim m_{23}$, the relation can be reduced to
\begin{align}
\text{sin}\,\theta_{13}\sim \frac{1}{2\sqrt{2}}\frac{m_{11}m_{33}}{m_{22}^2} ~\implies~ {m_{11}m_{33}}\sim 0.4 \times m_{22}^2~,
\end{align}
where \textit{the mixing angle $\theta_{13}$ is determined by the ratio between $m_{11}$, $m_{22}$, and $m_{33}$.} This relation is close to $m_{11}/m_{22}=m_{22}/m_{33}$, which might give us some hints about the generation of the Dirac masses and Yukawa couplings.

Finally, we can also estimate the value of the quasi-degenerate masses of $\nu_1$ and $\nu_2$ as
\begin{align}
m_1 \sim m_2 \sim \frac{m_{11}m_{33}}{M}\text{ sin}\,\theta_{23}
\sim m_3\text{ tan}\,\theta_{23}\text{ sin}\,\theta_{13}
\end{align}
which is $\sim 8.5$ meV taking $m_3=50$ meV. Some splitting between $m_1$ and $m_2$ as well as deviation from maximal mixing can be introduced if there is some cancellations among $m_{22}m_{23}$ and $m_{12}m_{33}$ in 23 entry or $m_{23}^2$ and $m_{13}m_{33}$ in 33 entry, which are natural as they are all of $O(\epsilon^2)$. In this case, $m_{\mu\mu}$ will be greater than $m_{\mu\tau}$ and $m_{\tau\tau}$, such deviation will lead to a splitting on the degenerate spectrum and an additional (but opposite) rotation to $\theta_{12}$, similar to the case (2) in section \ref{sec:IR2}. If such a deviation is small, as required by the observed large $\theta_{12}$, then one will predict a rather degenerate $m_1$ and $m_2$ with
\begin{align}
R_{21} \equiv \frac{m_2-m_1}{m_2+m_1}\sim 1-2\,\text{sin}^2\,\theta_{12} = 0.4 ~\implies~ \frac{m_2}{m_1}\sim 2.3 
\end{align}
Notice the analytical result above relies on the leading-term approximation, which might be interrupted if any subleading term proves to be comparable. Therefore, numerical study is required. Additionally, we will build a benchmark model based on our analysis.

\subsection{Numerical study and a benchmark model}

So far we have discussed the mass matrices analytically. In this subsection, we present a benchmark value for viable $m_D$ and $M_R$ which fit the current neutrino data. For numerical study, we consider the up-type quark masses at the high scale $\sim 10^9$ GeV \cite{Huang:2020hdv} for the input, which can remove the QCD effect and reflect a more natural relation among Dirac neutrino masses of different generations. To build a benchmark, we first take $m_{22}=m_{23}=400$ MeV as required from the maximal mixing, which also fixes the RH neutrino mass at the value of $M=5\times 10^9$ GeV. Next, we take $m_{33}=100$ GeV and all the rest of the entries are then fixed by the neutrino data. The two input matrices for eq.~\eqref{Lmass3} given by
\begin{align}
{m_D}\text{ (GeV)} =
\begin{pmatrix}
0.00058    &  0.00050    &  0.00023   \\
0.00050  &  0.4  &  0.4   \\
0.00023   & 0.4  &  100   \\
\end{pmatrix},~~
M_R\text{ (GeV)}=
\begin{pmatrix}
  0   &  0   &  5\times 10^9  \\
  0   &  -5\times 10^9 &  0  \\
  5\times 10^9    &  0   & 0  \\
\end{pmatrix}
\label{Mnum}\,.
\end{align}
The resulting LH neutrino mass matrix through the type-I seesaw mechanism is given by
 \begin{align}
{m_\nu} \text{ (meV) }=
\begin{pmatrix}
~\sim -10^{-6}    &&&   \sim -10^{-2}    &&&  -11.6~   \\
~\sim -10^{-2}  &&&  ~~~~~31.8  &&&  ~~21.8~   \\
~~~-11.6   &&& ~~~~~21.8  &&&  ~~22.8~   \\
\end{pmatrix}~.
\label{mnum}
\end{align}
The matrix results in the light neutrino mass spectrum given by
\begin{align}
m_1=7.4 \text{ meV},\quad m_2=11.4 \text{ meV},\quad m_3=50.7 \text{ meV}~,
\end{align}
which falls into the parameter space with a rather small $m_2/m_1$ region. The mixing angles and phases are given by
\begin{align}
\theta_{12}=35.35^{\circ},\quad \theta_{23}=49.1^{\circ},\quad \theta_{13}=8.54^{\circ},\quad \delta_{CP}=180^{\circ},\quad \eta_1=180^{\circ},\quad \eta_2=0^{\circ} ~,
\end{align}
where $\eta_1$ and $\eta_2$ are the Majorana phases of $m_1$ and $m_2$. The benchmark values are set to fit the PMNS matrix but with certain CP phases directly, except for the angle $\theta_{12}$, which is left free to make other parameters to get a better fit. The $\theta_{12}$ is chosen because the $U_{PMNS}$ is the combination of $U_\ell$ and $U_\nu$. Even if $U_\ell$ is similar to $U_{CKM}$ with small mixing, the $\theta_{12}$ of $U_\ell$ can still be sizable and help compensate the deviation we have in our benchmark. Compared to $\theta_{12}$, the other two angles are rather small in $U_{CKM}$ so these two angles in $U_{PMNS}$ are expected to be almost the same as the values from $U_\nu$.

The numerical study matches well with the analytical study in the previous subsection except for the ratio $m_2/m_1$, which is interrupted due to a non-negligible $\theta_{13}$, or more specifically $m_3\,$sin$^2\,\theta_{13}$. Notice that this benchmark model follows the assumptions on the Dirac mass matrix in eq.~\eqref{mD}, which are not necessarily true. It actually ends up with a mixing angle too small compared to the CKM matrix. Unless the CKM matrix is mainly from the down-type quark and charged lepton sector, which could be true as the observed hierarchy is also smaller compared to the up-type quark sector. Other assumptions can also be modified, including the relations $m_{22}m_{23}>m_{12}m_{33}$ and $m_{23}^2>m_{13}m_{33}$ among the $O(\epsilon^2)$ terms. Different conditions will require different substructures in the Dirac neutrino mass matrix, which are left for future study.


\section{Toward a UV complete model}\label{sec:UV}

A complete study for the dynamical generation of the seesaw scale through a strong interaction should also include the calculation of the $\beta$ function. The one-loop running coupling constant $g_H$ of $SU(2)_H$ is given by 
\begin{align}
\frac{d\,g_H}{d\,\text{ln}\,\mu}=-\frac{1}{16\pi^2}b_Hg_H^3 \quad\text{, where}\quad
b_H=\frac{22}{3}-\frac{1}{3}n_sT(R_s)-\frac{2}{3}n_fT(R_f)~.
\end{align}
The theory is only asymptotically free when $b_H$ is positive. It will then become strongly coupled and generate the condensate at the scale
\begin{align}
M_{\text{seesaw}}\sim e^{{-8\pi^2}/({b_Hg_0^2})}\cdot M_{\text{Planck}}~,
\end{align}
where $g_0$ is the coupling strength of $g_H$ at the Planck scale. Taking 
\begin{align}
M_{\text{seesaw}}\sim 10^9 ~\text{GeV} \quad\implies\quad b_Hg_0^2\sim 3.4
\end{align}
In the minimal case with only RH neutrinos as $3$ of $SU(2)_H$, we have $b_H=6$, which is already smaller than the $\beta$ function of $SU(3)_C$ in the SM gauge group. Therefore, a rather large $g_0\sim 0.75$ is required but is still a reasonable value.

However, the minimal case is not enough for a complete model. For a complete type-I seesaw mechanism, we need at least a Lagrangian given by
\begin{align}
-\mathcal{L}_{\text{seesaw}}=\overline{\ell_L}\,Y_\nu\,{H}\,\nu_R+\frac{1}{2}\,\overline{\nu_R^c}\,M_R\,\nu_R+\text{h.c.}~,
\label{Lseesaw}
\end{align}
which includes also the Dirac neutrino mass from the interaction with the Higgs field. If $\nu_R$ transforms as a triplet representation $\bf{3}$ under $SU(2)_H$, some part of the first term should also carry corresponding charges to make it $SU(2)_H$ invariant. As we already show, the $SU(2)_H$'s running is rather weak so we would like to minimize the additional fields charged under it. Making other fermions charged under $SU(2)_H$ is challenging since at least several of them need to be included to make the theory anomaly-free. Therefore, we expect other SM fermions to be charged under other horizontal symmetry and modify the scalar sector instead. We introduce a flavon field $F$ transforming as a $\bf{3}$ under $SU(2)_H$. To fit into the Yukawa term, it also needs to be a $\bf{3}$ under the horizontal symmetry of LH leptons. The flavon field is thus a $\bf{(3,3)}$ under two horizontal symmetries but a singlet under the SM gauge group. The Yukawa coupling thus originates from a higher dimensional operator given by
\begin{align}
-\mathcal{L}_{\text{Yukawa}}=\frac{1}{\Lambda_F}\overline{\ell_L}\, {H} \,F\,\nu_R+\text{h.c.}~,
\label{LYukawa}
\end{align}
where $\Lambda_F$ is the scale of new physics, such as the masses of exotic vector-like fermions. The nontrivial flavor structure of Yukawa coupling can then come from the vacuum structure of flavon fields, which should look like 
\begin{align}
\langle F_{33} \rangle   \sim  \Lambda_F ~,\quad 
\langle F_{22} \rangle  \sim  \langle F_{23} \rangle \sim \epsilon \,\Lambda_F ~,\quad
\langle F_{11} \rangle \sim  \langle F_{12} \rangle \sim  \langle F_{13} \rangle \sim \epsilon^2 \,\Lambda_F
\end{align}
to get the desired Dirac mass matrix. The mystery of the Yukawa couplings can now be traced back to the vacuum structure of the flavon. The discussions on the flavon potential and VEV are beyond the scope of this work so we leave them for future study. With only a $\bf{(3,3)}$ flavon field, which includes three $SU(2)_H$ triplet scalar fields, we can still get a positive $b_H=4$ with a reasonable coupling $g_0\sim 0.9$.

Notice that the VEV of the flavon field $\langle F \rangle$, which is a triplet under $SU(2)_H$, will also break the $SU(2)_H$ gauge symmetry and make it a broken, unconfined strongly coupled theory like the Nambu-Jona-Lasinio (NJL) model \cite{Nambu:1961tp, Nambu:1961fr}. In this case, the RH neutrinos that participate in the strong interaction can still be freely propagating degrees of freedom below the condensate scale, which is the seesaw scale $M_{\text{seesaw}}$ in our case. We can then get rid of dangerous consequences from the confinement, similar to the top quark condensation models \cite{Miransky:1988xi, Miransky:1989ds, Marciano:1989xd, Bardeen:1989ds, Hill:1991at} where top quarks undergo a non-confining strong interaction.


\section{Testing the idea: from neutrino data to gravitational waves}\label{sec:Test}

In this section, we list all the possible phenomenological consequences of our model all the way from the light LH neutrinos at the milli-eV scale (the low-energy neutrino data) to the heavy RH neutrinos and strong dynamics at the exa-eV scale (gravitational waves).

Start from low-energy neutrino experiments. As we already discussed in section \ref{sec:3G}, the right-handed neutrino mass matrix with assumptions on the Dirac neutrino mass matrix leads to a special flavor structure on the light neutrino mass matrix $m_\nu$ shown in eq.~\eqref{mnu}, which, at the leading-order approximation, is the same as the $A_1$ mass matrix with two-zero textures \cite{Frampton:2002yf}. The phenomenology of neutrino mass matrix with $A_1$ texture have been well studied in \cite{Fritzsch:2011qv,Zhou:2015qua,Alcaide:2018vni}, which includes several important features as follows:
\begin{align}
\text{(1) Normal Ordering}\qquad \text{(2) Large $\delta$ (between $180^{\circ}$ and $270^{\circ}$)}\qquad  \text{(3) Small $m_{\beta\beta}$}~.\nonumber
\end{align}
The first two properties will be tested by the next-generation long-baseline neutrino oscillation experiments, such as Hyper-Kamiokande \cite{Hyper-Kamiokande:2018ofw}, DUNE \cite{DUNE:2016hlj}, and JUNO \cite{JUNO:2015zny}, in the near future. On the other hand, the third property predicts a null result for the neutrinoless double beta decay even in the next-generation experiments \cite{LEGEND:2021bnm, nEXO:2021ujk}. However, it also means any signal observed could put a strong constraint on our idea.

Next, restricting to the one-parameter $M_R$ matrix, the substructure of the Dirac neutrino mass matrix is determined by the observed neutrino spectrum and mixing. Therefore, if the origin of Yukawa coupling is somewhat within reach, such as at the TeV scale, and can be measured in the future, the matching between neutrino mixing and the Dirac mass matrix, such as two relations we derived
\begin{align}
(1)~ \text{tan}\,\theta_{23} \sim \frac{m_{22}}{m_{23}}\ ,~\quad~
(2)~ \text{sin}\,\theta_{13} \sim \frac{m_{11}m_{33}}{m_{22}^2+m_{23}^2}\frac{m_{23}}{\sqrt{m_{22}^2+m_{23}^2}} ~,
\end{align}
can then be tested. However, if the measurement is through the quark sector, it will then depend on additional assumptions of the quark-lepton relation.

The first two approaches, focusing on either the light neutrino mass matrix or the Dirac mass matrix, can only provide indirect probes of our idea, given that our main mechanism is the strong dynamics among the RH neutrinos. The direct test of the idea thus requires access to the seesaw scale, where the condensation occurs. The strong dynamics introduce three degenerate RH Majorana neutrinos with $M\sim 10^9$ GeV, which are hard to access. One common approach is through the studies of leptogenesis \cite{Fukugita:1986hr}. However, such a thermal process also relies on several assumptions and the degenerate spectrum of RH neutrinos in our model seems unlikely to lead to a successful thermal leptogenesis due to the strong washout effects \cite{Akhmedov:2003dg}.

Besides the RH neutrinos themselves, bound states of RH neutrinos are also formed, which might have impacts on the phenomenology. However, most of the bound states also sit at the seesaw scale, except the majoron \cite{Chikashige:1980ui}, which is the Goldstone boson of broken $U(1)_R$ symmetry. The properties of majoron could provide some new information about the origin of neutrino masses. If the $U(1)_R$ global symmetry is exact, the majoron will be massless, which is then accessible at low energy. However, this is usually not the case. For example, if $U(1)_R$ is gauged, the Goldstone mode will be eaten and we get a seesaw-scale gauge boson instead, again inaccessible. Additional interactions that don't follow the global symmetry will also introduce a nontrivial potential for the majoron, which could come from, for example, the neutrino Yukawa coupling term in eq.~\eqref{LYukawa} if other fields don't carry $U(1)_R$ charges.

Although the new degrees of freedom might only exist at the seesaw scale well beyond the reach of colliders, the mechanism can still leave an imprint on the universe. Since the first observation of gravitational waves, a novel approach to explore particle physics at a high scale is now available, with many dedicated studies at probing the seesaw scale \cite{Okada:2018xdh,Brdar:2018num,Dror:2019syi,Fu:2023nrn}. Such a signal in principle can distinguish between different dynamics, which can thus provide important inputs for the neutrino mass and flavor model building at the seesaw scale. The gravitational waves from the phase transition of strongly coupled theories have been explored in \cite{Jarvinen:2009mh,Schwaller:2015tja,Chen:2017cyc,Helmboldt:2019pan,Agashe:2019lhy,Halverson:2020xpg,Huang:2020crf,Kang:2021epo,Reichert:2021cvs,Pasechnik:2023hwv}, revealing a distinct pattern compared to those derived from weakly coupled theories. However, there are two concerns when using this new approach to test our idea.

First, to generate gravitational waves, the phase transition needs to be strong first order, but the $SU(2)$ phase transition is second order \cite{Svetitsky:1982gs,Svetitsky:1985ye,Kogut:1982fn,Kogut:1982rt}. Unless the strong $SU(2)$ in our model comes directly from a strong $SU(3)$ through the tumbling mechanism \cite{Raby:1979my}, which is possible for a $SU(3)$ chiral gauge theory with a sextet representation $\bf{6}$ \cite{Amati:1980wx,Gusynin:1982kp}. The fermion in the two-index symmetric representation can also increase the strength of the first-order confinement phase transition \cite{Reichert:2021cvs,Pasechnik:2023hwv}. However, the second problem is that, even with a sufficiently strong signal, the highest scale a detector can explore is limited by the upper frequency it can reach. The currently planned gravitational wave detectors can only reach the upper frequency at the order of $10^3$ Hz, which corresponds to the upper bound on the detectable energy scale at the order of $10^8$ GeV, still below our proposed seesaw scale. Therefore, testing our idea through gravitational waves will require both a more complicated strong sector and new technology for future observatories.


\section{Conclusions and outlook}\label{sec:Conclusion}


We examine the idea of using strong dynamics to realize the type-I seesaw mechanism while ensuring consistency with the low-energy neutrino data. The RH neutrino Majorana mass with a nontrivial flavor structure is dynamically generated at the seesaw scale. Assuming a Dirac neutrino mass matrix with hierarchical eigenvalues and small mixing angles, analogous to the up-type quark sector, we find that the desired strongly anti-diagonal RH neutrino mass matrix $M_R$ can be generated from a strong $SU(2)_H$ horizontal gauge symmetry with RH neutrinos transforming as a triplet representation $\bf{3}$. In this strongly coupled theory, the seesaw scale emerges when the coupling $g_H$ becomes strong in the infrared, which explains the origin of the seesaw scale. Additionally, a nontrivial flavor structure for the mass matrix $M_R$ is introduced following the criterion of the strong horizontal gauge interaction.


The strongly anti-diagonal flavor structure of $M_R$ can enhance the substructures inside the Dirac mass matrix $m_D$, which results in the observed large mixing and weak hierarchy in the neutrino flavor structure. These substructures, if exist, might provide a novel direction to probe the unknown rules behind the Yukawa couplings. They are smoothed out due to the strong hierarchical diagonal entries when looking into the CKM matrix of the quark sector but are now reflected in the PMNS matrix of the lepton sector thanks to the seesaw enhancement. In this work, we reveal these substructures through a bottom-up approach with many assumptions for simplicity. Modification of the assumptions can lead to other possibilities, which are left for future study. Furthermore, if one can derive these substructures from a top-down approach, it could provide an economical UV-complete model for the flavor puzzles, warranting further investigation.


Besides achieving the desired large mixing and weak hierarchy, the resulting light neutrino mass matrix $m_\nu$ is actually over-constrained, providing a unique flavor structure. The structure, similar to the $A_1$ matrix with two-zero textures, has several phenomenological consequences including the spectrum in a normal ordering and the range of the CP-violating phase, which can be tested in the near future. The small $m_{\beta\beta}$ also predicts a null result for the neutrinoless double beta decay. The direct test of the idea will require access to the seesaw scale such as gravitational waves, which can probe the dynamics happening in the early universe. However, the signal range needs to be further expanded in the future to cover the most relevant scale. If feasible, the detection of a specific spectrum from strong dynamics at the predicted seesaw scale could serve as a compelling evidence for our mechanism.

\acknowledgments

I thank Manfred Lindner, Alexei Y. Smirnov, and \'Alvaro Pastor-Guti\'errez for many useful suggestions and discussions.


\bibliographystyle{jhepbst}
\bibliography{Neutrino_Ref}{}

\end{document}